# Hydrogen loss and its improved retention in hydrogen plasma treated *a*-SiN$_x$:H films: ERDA study with 100 MeV Ag$^{7+}$ ions

R K Bommali[1], S Ghosh[2], S A Khan[3] and P Srivastava[2]

[1]Institute of Physics Bhubaneswar, Sachivalaya Marg, Bhubaneswar, Odisha 751005, India [2]
[2] Nanostech Laboratory, Department of Physics, Indian Institute of Technology Delhi, Hauz-Khas, New Delhi 110 016, India
[3]Materials Science Division, Inter-University Accelerator Centre (IUAC), Aruna Asaf Ali Marg, New Delhi 110067, India

Email: ravibommali@iopb.res.in

**Abstract.** Hydrogen loss from *a*-SiN$_x$:H films under irradiation with 100 MeV Ag$^{7+}$ ions using elastic recoil detection analysis (ERDA) experiments is reported. The results are explained under the basic assumptions of the molecular recombination model. The ERDA hydrogen counts are composed of two distinct hydrogen desorption processes, limited by rapid molecular diffusion in the initial stages of irradiation, and as the fluence progresses a slow process limited by diffusion of atomic hydrogen takes over. Which of the aforesaid processes dominates, is determined by the continuously evolving Hydrogen concentration within the films. The ERDA measurements were also carried out for films treated with low temperature (300ºC) hydrogen plasma annealing (HPA). The HPA treated films show an improved diffusion of atomic hydrogen, resulting from healing of weak bonds and passivation of dangling bonds. Further, upon HPA, films also show evolution of hydrogen with significantly higher counts, at advanced fluences, relative to the as-deposited films. These results indicate the potential of HPA towards improved H retention in *a*-SiN$_x$:H films. The study distinguishes clearly the presence of two diffusion processes in *a*-SiN$_x$:H whose diffusion rates differ by an order of magnitude, with hydrogen radicals not being able to diffuse beyond ~1 nm from the point of their creation. The results are very relevant for the passivation applications of *a*-SiN$_x$:H.

## 1. Introduction

Sub stoichiometric silicon nitride thin films deposited by plasma enhanced chemical vapour deposition technique (PECVD) find wide range of applications like device passivation [1], LEDs [2],

ARCs [3,4] for solar cells [5] and devices [6]. Due to the use of precursor gases like silane ($SiH_4$) and ammonia ($NH_3$) these films have an inevitable presence of hydrogen in them, which leads to the popular nomenclature of 'PECVD silicon nitride' as 'amorphous hydrogenated silicon nitride' and abbreviated as '$a$-SiN$_x$:H'. The process induced hydrogen incorporated in these films plays a crucial role in its microstructural [7], electronic [8] and optical properties [9,10]. Hydrogen effusion from $a$-SiN$_x$:H is relevant to many post processing steps like UV exposure [11], high temp annealing [12], rapid thermal annealing [13-15], contact firing step in solar cell passivation [16,17], swift heavy ion irradiation, EUV irradiation [18]. In the past the study of loss of hydrogen from materials like $a$-C:H [19], $a$-Si:H [20], $a$-Si:C:H [21] and organic polymers [22] have been reported. In these reports, hydrogen loss has been widely accepted to take place predominantly in molecular form rather than in the atomic form. The release basically follows a single exponential decay or a combination of them depending on the microstructure and composition of the matrix hosting hydrogen. Accordingly, various models [23] to suit the particular case of these materials have been suggested.

Presently, we investigate the hydrogen release processes for the specific case of as-deposited and hydrogenated $a$-SiN$_x$:H thin films during elastic recoil detection analysis [24] (ERDA) measurements employing 100 MeV Ag$^{7+}$ beam. It must be noted that irradiation with energetic (~MeV) ions offers a controlled and reproducible means of effecting hydrogen release from materials under extreme conditions, wherein, energy deposition can be varied in a wide range by the choice of the ion and its energy, in contrast to the conventional thermal treatments. The energy deposition of the ions into $a$-SiN$_x$:H thin films are understood on the basis of thermal spike model. The present article discusses various factors that determine the release of hydrogen during MeV ion irradiation.

In the past, hydrogen release from $a$-SiN$_x$:H deposited by various growth methods like hotwire CVD [25], ECR Plasma deposition [14], reactive evaporation [26] and PECVD [12] have been reported. These reports have employed various thermal treatments like RTA or furnace anneals, which are quite distinct from the ion irradiation process. However, the detailed study of the effect of stoichiometry on effusion of hydrogen from the $a$-SiN$_x$:H films during ERDA with MeV ion irradiation is being reported for the first time to the best of our knowledge. Furthermore, the present study is also extended to films subjected to low temperature hydrogen plasma annealing (HPA), wherein, an order of magnitude improvement in H retention under MeV irradiation is observed. This improvement is attributed to the remarkable healing of the $a$-SiN$_x$:H network during low temperature HPA.

## 2. Experimental details

SRSN thin films were deposited on p-type Si (100), by conventional PECVD (Model: SAMCO PD-2S) technique. This system uses RF (13.56 MHz) power to maintain a glow discharge and produce ionized species necessary for thin film growth. Silane ($SiH_4$, 4% in Ar) and ammonia ($NH_3$) were used as precursor gases. The partial pressure ratio $P_r = P_{SiH_4}/P_{NH_3}$ of the reactant gases was varied to deposit films of different stoichiometries, which are named as S1, S2, S3, S4 and S5 hereafter. The Si content decreases from film S1 to S5. The other deposition parameters namely substrate temperature, chamber pressure and plasma power were maintained at 200°C, 1 mbar and 25 W respectively. Further, the deposition time was varied in order to limit the thickness of the SRSN layers to ~100 nm. Low-temperature HPA was carried out for one hour at three different temperatures (250°C, 300°C and 350°C) in the PECVD system. During HPA, the pressure, RF power and hydrogen flow were kept at 2 mbar, 25 W and 10 sccm respectively. Rutherford Backscattering Spectrometry (RBS) experiments were carried out using 1.7 MeV $He^{2+}$ ions at a backscattering angle of 170°. The detector solid angle was 3.052 milli-steradians and its resolution was 17 keV. The concentration (atoms/nm$^3$) of hydrogen in both as-deposited and HPA films was measured by the ERDA technique using 100 MeV $Ag^{7+}$ ions. ERDA experiments were carried out in a vacuum of 4.5×10$^{-6}$ mbar and a collimated beam of silver ions of a spot size of 1×1 mm$^2$ was made to impact at an angle of 20° with respect to the film surface. The hydrogen recoils from the films were detected in a silicon surface barrier detector (SSBD) kept at 30° recoil angle with a polypropylene stopper foil in front of it to stop other recoils like nitrogen (N), oxygen (O) and silicon (Si).

Table I. Details of the samples used for ERDA study

| Sample | Si | N | excess Si atomic % | Si$_3$N$_4$ %age | N/Si | growth rate (nm/min) | Thickness (nm) ± 0.1 | Density* gm/cc |
|---|---|---|---|---|---|---|---|---|
| S1 | 0.45 | 0.15 | 74.9 | 25.1 | 0.3 | 55 | 109.6 | 1.72 |
| S2 | 0.4 | 0.2 | 62.4 | 37.6 | 0.5 | 54 | 117.8 | 1.61 |
| S3 | 0.3 | 0.3 | 24.8 | 75.2 | 1.0 | 42 | 91.7 | 1.73 |
| S4 | 0.31 | 0.34 | 17.5 | 82.5 | 1.1 | 40 | 116.8 | 1.90 |
| S5 | 0.24 | 0.3 | 6.0 | 94.0 | 1.3 | 28 | 127.5 | 1.98 |

*Film densities as determined from the critical angle obtained in X-ray reflectivity measurements

## 3. Results and Discussion

Table I lists the various details of the samples used for the ERDA experiments. It should be noted that all the samples have comparable thickness. Figure 1 shows the results of ERDA measurement on the $a$-SiN$_x$:H thin films labelled S1 to S5, wherein there is a change in stoichiometry going from S1 to S5 (see Table I). The data basically represents the counts of hydrogen recoils (atoms/cm$^2$) from $a$-SiN$_x$:H films upon irradiation with 100 MeV $Ag^{7+}$ ions.

The obtained data represents the counts from the ERDA detector are normalized according to the thicknesses (see Table I) of the films in order to obtain the Hydrogen concentration per unit volume. There are several points to be noticed. Firstly, it may be noted that all the decay curves (plotted on a logarithmic y axis) are seen to be composed of two linear regions thereby indicating the presence of two hydrogen loss regimes. A fast mechanism dominates at fluence below ~10$^{12}$ ions/cm$^2$ whereas a slow release dominates at higher fluences. ERDA data for hydrogen loss can be fit with a double exponential describing the aforementioned decays:

$$N(\emptyset) = y_o + N_1 e^{-K_1 \emptyset} + N_2 e^{-K_2 \emptyset} \qquad (1)$$

$N(\emptyset)$ is the hydrogen counts at any given fluence ($\emptyset$), $y_o$ represents the background noise. $N_1$ and $N_2$ are the H concentrations associated with each of the evolution processes. The sum, $N_1+N_2$ gives the initial hydrogen content of the films prior to irradiation. $K_1$ & $K_2$ represent the rate constants for the two decay mechanisms. Secondly, the decay rates for these apparently independent mechanisms undergo change as the film composition varies from Si rich to stoichiometric. This indicates clearly a difference in diffusion constant of H as the films composition changes. We will attempt to understand the dependence of the hydrogen diffusivity through *a*-SiN$_x$:H films as the films composition (N/Si) on one extreme (S1) approaches *a*-Si and silicon nitride (S5) on the other end, in later sections of this article to understand.

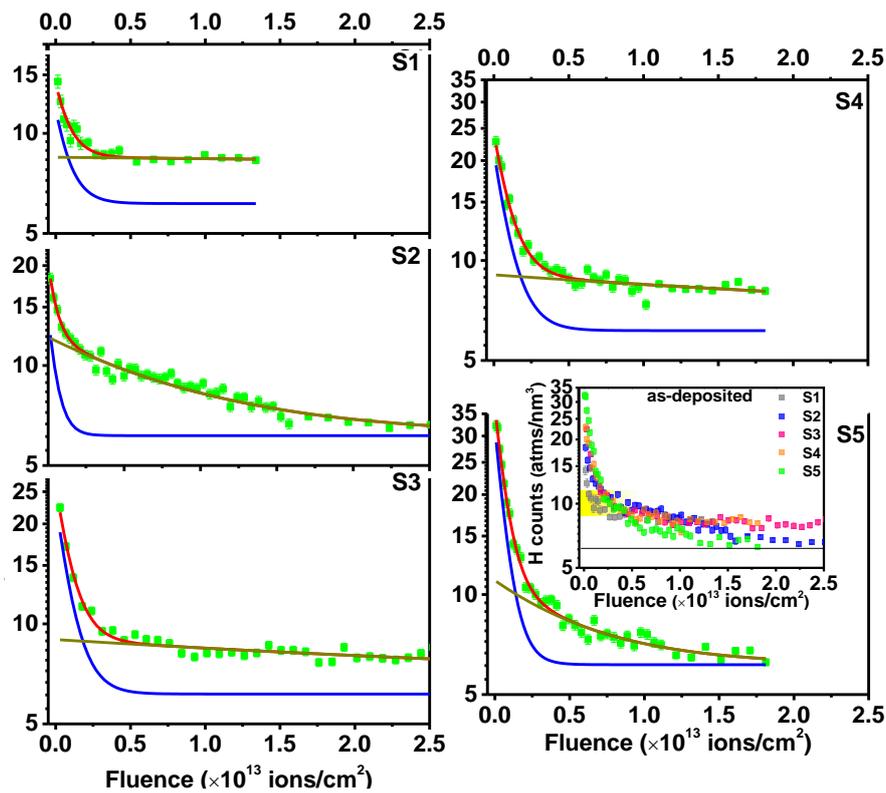

**Figure 1.** Hydrogen counts as measured in the ERDA experiment for as-deposited *a*-SiN$_x$:H thin films S1 to S5 and their fitting with two exponentials decays, which represent different hydrogen release mechanisms. Inset shows the data overlapped on one another, wherein the yellow strip indicates the H concentration below which the H release slows down significantly.

Further, we will also investigate the conditions that determine the passage from one decay process to the other. The factors responsible for the asymptotic decay of H content to high fluences, following the initial rapid depletion will also be discussed.

*3.1. Ion matter interactions in a-SiN$_x$:H*

In order to understand the presently obtained data let us first take a brief survey of the current understanding of the hydrogen release processes. In general, the thermal/UV/MeV-ion based excitation of the hydrogenated materials leads to extensive bond breaking and restructuring inside these materials. For the present study of a-SiN$_x$:H we only consider the hydrogen related bonds. The Hydrogen related bonds viz. Si—H and N—H are broken leading to the generation of atomic hydrogen/radicals. The H radicals so generated recombine to form H$_2$ molecules, as and when they come within close vicinity of each other. These H$_2$ molecules diffuse through the material to escape out of the material surface. Apart from this aforementioned sequence, in their transit through the material the radicals can also abstract a hydrogen atom bonded to Si or N and form a H$_2$ molecule. Concurrently, the abstraction process also leads to formation of a Si—N or Si—Si or N—N bonds based on the immediate vicinity of the bond from which the bonded hydrogen was abstracted. Each of the bonds has different formation energies [12,14] and therefore the kinetics of the individual processes may vary. In addition to these release mechanisms there are sites like Si or nitrogen dangling bonds that may trap the H radicals, thereby limiting their diffusion to small distances from the sites of their creation. In comparison, H$_2$ molecules are inert and thereby diffuse relatively fast and unhindered through the matrix.

Coming to the specifics of the present case, in the ion matter interactions, the energy deposition in the material is quite different from the other processes and needs further description. The interaction of swift heavy ions with matter is currently best explained under the thermal spike formalism [27]. Energetic (few 10's to 100 MeV) ions during their passage through matter, interact predominantly through their electron cloud. The ions kinetic energy is in-elastically deposited into the material through the electronic subsystem to the phonon system of the material leading to heating of the material. Since the ion passes at very high speeds through the material the deposited energy is extremely short lived (~fs) and localized. As it spreads radially outwards from the ion trajectory as its axis, it forms a cylindrical volume of damaged/affected material known as the ion track. The localized nature of the energy deposition process also means that the volume beyond a certain ion track can be considered unaffected by the temperature rise and the ensuing damage within it. The damage can be thus assumed to be built up cumulatively with individual ion passage events. It is only after a certain value of fluence that the complete coverage of the sample volume is achieved. Going beyond this threshold fluence the ions are assumed to overlap and accumulate further damage on the now non-pristine volume. The threshold fluence at which the tracks overlap is called the track overlap fluence, and essentially depends on the ion track diameter, which in turn is determined by the electronic energy loss ($S_e$) for a given ion-material combination. In the case of hydrogen loss by MeV ions the overlap fluence is marked by a sharp change in the rate of H loss values. For example, for the currently obtained data the 'overlap fluence' lies in the region ($\sim 1\times 10^{12}$ to $5\times 10^{12}$ ions/cm$^2$) where the slow exponential decay begins to dominate over the initial fast decay. This decrease can be partly attributed to the drastic depletion of the H concentration during the pre-overlap regime. However, an order of magnitude difference in decay rates in the pre and post overlap regimes points out to the quite distinct mechanisms at work in the two regimes. Identification of these mechanisms in the particular case of $a$-SiN$_x$:H is endeavoured hence.

*3.2. Hydrogen release in a-SiNx:H*

In $a$-SiN$_x$:H hydrogen is incorporated primarily as Si—H and N—H bonds. Further, the presence of molecular hydrogen [28] in the interstitials and voids cannot be ruled out. The passage of MeV ions can lead to an almost instantaneous increase in temperatures to the order of few ~1000 K, sufficient to cause local melting of the material. Thus deposited energy is well beyond sufficient to disrupt hydrogen related bonds, eventually leading to the formation of H$_2$ molecules which diffuse out of the material with a rate proportional to the concentration of molecular hydrogen in the films. As estimated from ERDA counts the initial H concentration of the as-deposited films varies between ~15 to 35 atoms/nm$^3$ as the nitrogen content increases going from sample S1 to S5. The number compares very well with the number of Si and N atoms per unit volume, which are ~ 24 and 32 atoms per nm$^3$ respectively, for a PECVD nitride of density ~2 gm/cc. This means that subsequent to the disruption of bonds the H radicals have to hop only a few atomic sites before combining with one another, thereby leading to the formation of molecular hydrogen almost instantaneously. Thus in the initial stages of irradiation the hydrogen loss is limited by diffusion of molecular hydrogen. However, with continued irradiation and rapidly progressing depletion of the H concentration, the overall H counts decay rates tend to decrease. In order to understand this transition into the slow decay regime we must first notice that this onset occurs, for all films in consideration, when the H recoils falls close to critical concentration ($H_{critical}$) of ~8-10 atoms/nm$^3$, indicated by yellow band in the inset of figure 1. This then indicates that as the H recoils falls below $H_{critical}$, recombination of H radicals to form a $H_2$ molecule becomes increasingly difficult. In this region the release of $H_2$ is limited by the availability of H radical pairs in close vicinity, within a few atomic spacings. The low probability of the recombination is further aggravated due to the fact that these recombinations must occur in the short time scale when the ion track is 'hot' and has sufficient energy to support atomic or molecular diffusion of hydrogen. As a consequence only radicals formed within a given ion track recombine to form molecules. This will hold true unless tracks are formed simultaneously in the immediate proximity of each other. The schematic representation of the hydrogen effusion is summarized in figure 2. The aforementioned conclusions also mean that if the H concentration/density is sufficiently depleted (~10 atoms/nm$^3$) to begin with then we can only expect the slower (single exponential) release process limited by Hydrogen radical diffusion to be present in the ERDA data. This above said expectation is evidenced in the work of Singh *et al* [29], where ERDA was carried out on as deposited films and swift heavy ion irradiated films and a comparison was made. It was observed that for films subjected to high fluence irradiation (with 100 MeV $Ni^{7+}$ ions at $1\times10^{14}$ ions/cm$^2$) the ERDA hydrogen loss curve comprises of single decay mechanism. In contrast, the hydrogen loss obtained for as-deposited films exhibited both the decay processes. This observation confirms the understanding of hydrogen release presented above.

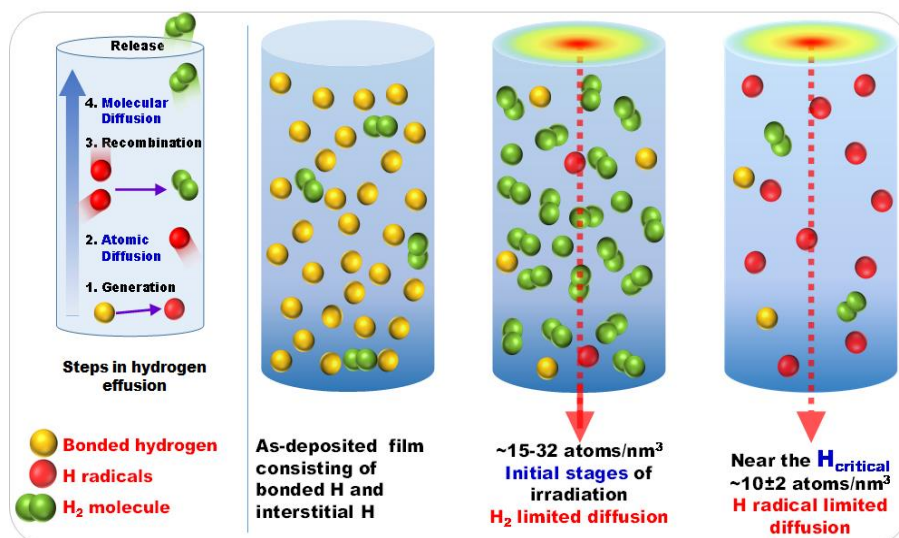

Figure 2. Schematic of hydrogen release from $a$-SiN$_x$:H under MeV ion irradiation.

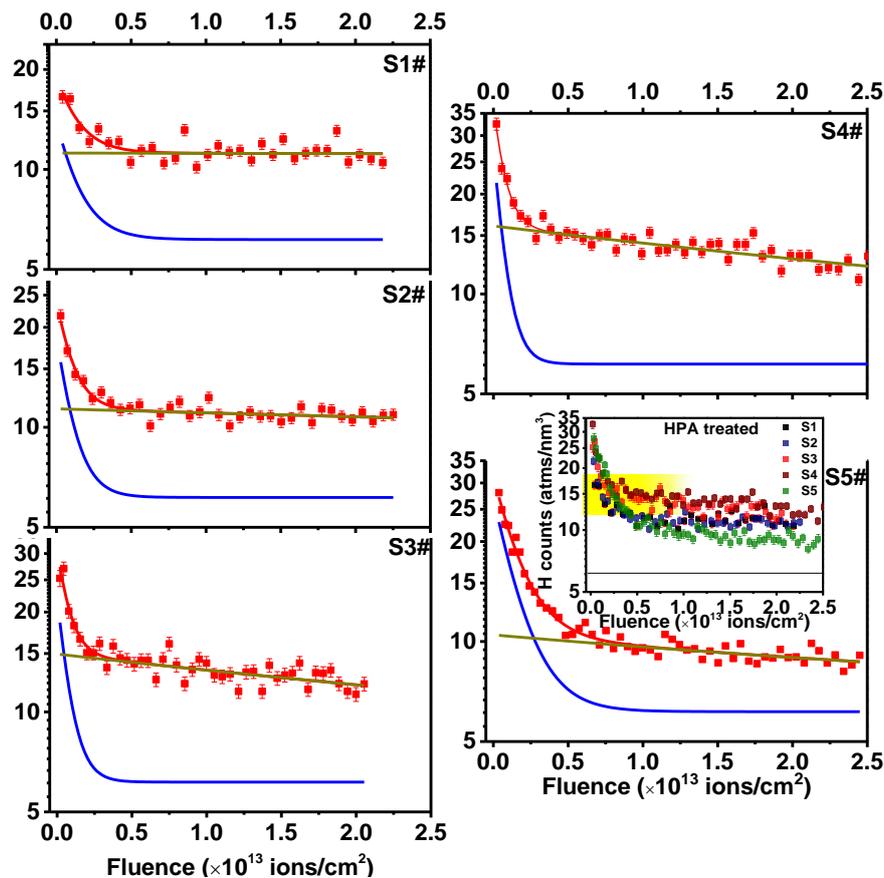

**Figure 3.** The H release data for the post-hydrogenated films. Notice the onset of the slow decay regime at ~12-16 atms/nm$^3$ as indicated by the yellow strip.

*3.3. Effect of hydrogen plasma*

Further, hydrogen plasma treatments were carried out to look out for possible changes in H release mechanisms upon hydrogenation of the films. Figure 3 shows the ERDA data for the *a*-SiN$_x$:H films treated with low temperature 300ºC hydrogen plasma for 1 hour. There are a few interesting points that can be quickly noticed from the data, in comparison to the as-deposited films (figure 1), regarding the effect of the hydrogenation:

- An early onset of the slow decay regime at $H_{critical}$ of ~14 atoms/nm$^3$ as compared to ~8 atoms/nm$^3$ the as-deposited films.
- The persistence of higher hydrogen counts towards advanced fluences ($>1\times10^{12}$ ions/cm$^2$).
- Increase in the initial H concentration of films.

Firstly, it may be noticed that the onset of the atomic diffusion process is drastically shifted to higher $H_{critical}$ (~14 atoms/nm$^3$). This increase in $H_{critical}$ indicates towards an improved diffusion of atomic hydrogen in the films compared to the as-deposited film. The passivation of dangling bonds and curing of weak or strained bonds with an overall improvement in the amorphous *a*-Si:H network by means of, a disorder to order transition; an established result as studied and reported extensively in the past [30]. In the present case the aforesaid passivation effect leads to improved diffusion of atomic radicals through the *a*-SiN$_x$:H network. The passivation of dangling bond defects during hydrogen plasma treatments as also evidenced in our previous work on the enhancement of photoluminescence yield upon plasma hydrogenation [31]. Further, the improvement in the silicon nitride network concomitant to the passivation is clearly visible from the significantly enhanced counts of hydrogen recoils observed at fluence exceeding ($>1\times10^{12}$ ions/cm$^2$) for the HPA treated films. The aforesaid observation clearly indicates the improved hydrogen retention in the HPA treated films under irradiation. In order to visualize the post HPA improvement in the hydrogen retention of the films, we determine the fluence at which the hydrogen concentration of the films drops to 10% of the initial hydrogen content and call it depletion fluence ($\phi_d$). Figure 4a shows a graph of $\phi_d$ vs the N/Si ratio. It is seen that Si rich films endure better (by ~2 orders) against irradiation induced hydrogen depletion. This in other words is an indicator of the poor mobility of H radicals as the films approach *a*-Si:H stoichiometry [20]. Further, the HPA treatments are found to improve the H retention by an order of magnitude through all the N/Si ratios considered here. This result is the clear evidence of the repair and a consequent improvement in the silicon nitride network post HPA treatment. In contrast the HPA effected improvement in photoluminescence showed a maximum enhancement for intermediate stoichiometries ($x$~0.8) [32]. This difference is interesting and can be a subject of future investigations. Another consequence of the HPA treatment is the increase in the H content of the films. This is made clear from the total initial hydrogen concentration $N_1+N_2$ (see equation 1), as determined from the fits of the ERDA data for the as-deposited and the hydrogenated films, depicted in figure 4b. Hydrogen gets incorporated in the *a*-SiN$_x$:H network by attaching on to dangling bonds as well as in the form of molecular hydrogen at interstitial sites. It must be noted herein that it is difficult to infer directly from the ERDA data about the nature of hydrogen incorporation into the *a*-SiN$_x$:H films.

*3.4. Rates of hydrogen desorption*

With the fundamental understanding of the post and pre-overlap decay mechanisms, discussed in the previous sections, the factors that determine the decay rates of H depletion can be discussed. It must be noted that the decay rates observed in both the regimes show a dependence on stoichiometry. In the pre overlap regime, the H concentration is high, so that upon ion impingement H radicals are readily formed from the Si—H and N—H as well as from interstitial $H_2$ molecules. The formation rate of these radicals can be safely considered to be instantaneous compared to the decay time scales considered here, given the extremely short-lived energy deposition by the impinging ion. Further, since we assumed a high H content in the material for this region, it is likely that the radicals recombine relatively fast to give rise to a surge of molecular hydrogen. This surplus $H_2$ then tries to equilibrate inside the film and then eventually be released out of the film surface due to the concentration gradient. In these circumstances it is plausible to conclude that the decay rate in the pre-overlap regime is limited and determined by the diffusion of the $H_2$ molecules through the films. In turn, factors like voids, dangling bonds and the initial hydrogen content determine the $H_2$ diffusion. For example, higher void fraction promotes $H_2$ diffusion on the other hand dangling bonds may inhibit the $H_2$ diffusion. Further, higher initial H content will promote a higher rate of depletion, as dictated by the underlying process of diffusion. These factors are determined by the composition and growth rate of the films. On the other hand, the slow decay regime, as discussed before, is limited by the H radical diffusion through the material. The H radical, unlike the $H_2$ molecule, has a slow passage through the material owing to its chemically excited nature. It may participate in H abstraction, from Si—H and N—H bonds or it may recombine with other radicals. Both these processes lead to the formation of a $H_2$ molecule before its eventual release from the material. It may also happen that the H radical is trapped by a Si or N dangling bond defects before the aforesaid recombination. We must also take note of the other concomitant modifications in the $a$-SiN$_x$:H microstructure in the high fluence ($>5\times10^{12}$ ions/cm$^2$) regime. The onset of the high fluence regime is marked by extensive bond breaking and restructuring of the Si—N network, leading to the collapse of the void structures within the films. For example, for films irradiated upto ~$1\times10^{14}$ ions/cm$^2$ the film thicknesses are found to reduce by 10-15% [29]. It has been evidenced in X-ray reflectivity measurements (not shown here) that significant densification and compaction of the films takes place only after the overlap fluence ~$1\times10^{12}$ ions/cm$^2$. This restructuring, because of the short time scale of the deposited energy, leads to the formation of dangling bond defects. For example, with the progression of irradiation fluence the formation of Photoluminescence quenching non-radiative defects was evidenced in our previous work [30]. The reduction in the voids and the extensive formation of dangling bonds in the high fluence regime further impedes the migration and a resulting recombination of H radicals. Therefore we conclude that the compositional dependence of the decay rates in the post overlap regime is determined by the extensive microstructural modifications that evolve in the films depending on their stoichiometry. We now look at the decay rates in the two regimes by the double exponential fits, as shown in figure 1 and figure 3, of the experimental data with equation 1. Figure 5a and 5b show the variation in decay constants, $K_1$ and $K_2$ respectively, for as-deposited and HPA treated films with composition (N/Si ratio). Firstly, it may be observed that the initial fast decay ($K_1 \sim 10^{-13}$) is an order of magnitude faster than the slow release ($K_2 \sim 10^{-14}$) that occurs at higher fluence. The green curves in figure 4a and 4b represent $K_1$ and $K_2$ for the as deposited films. We can notice a similar S shaped dependence for the both the curves, which indicates a common origin of the factors responsible for the two hydrogen mechanisms, vis-à-vis, the void percentage and the dangling bond concentration. Also plotted in figure 5c, is the variation of the density of the as-deposited films as determined from the critical angle in X-ray reflectivity data. It may be noted that the film density has a trend (S shaped) comparable to $K_1$ and $K_2$ of the as-deposited films. This points towards the role of film density, convoluted with factors like defects on these voids surfaces, in determining the decay rates observed herein. Further coming to the decay rates after HPA (red curves), trends for both $K_1$ and $K_2$ modify from an S shaped variation into a linear increase with N/Si ratios. Only deviation in the trend is for S5, wherein the post HPA $K_1$ value falls below the linear trend. This may be attributed to the post HPA decrease in the H content of the films as can be observed from figure 4. It must be noted that according to the theoretical work of Sun *et al* on the UV illumination induced H release, the probability of molecular hydrogen formation increases as the density N—H and Si—H bonds tend to balance as the Nitrogen content is increased [33]. This points out to the drastically different conditions prevailing in ion matter interaction processes and hence the requirement of a separate treatment for the hydrogen release by MeV ions. However, the confirmation of these correlations between dangling bond density and the void fraction on the diffusion of atomic and molecular hydrogen is beyond the scope of the present work and would require further detailed experiments.

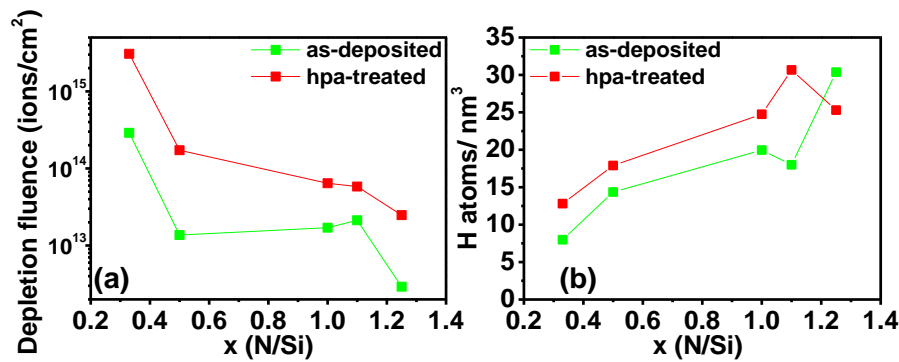

Figure 4. (a) The "depletion fluence" required to attain a hydrogen concentration of 10% of the initial value, plotted vs the N/Si ratio, before and after hydrogenation of the films. (b) The total initial hydrogen content for the as-deposited and hydrogenated films

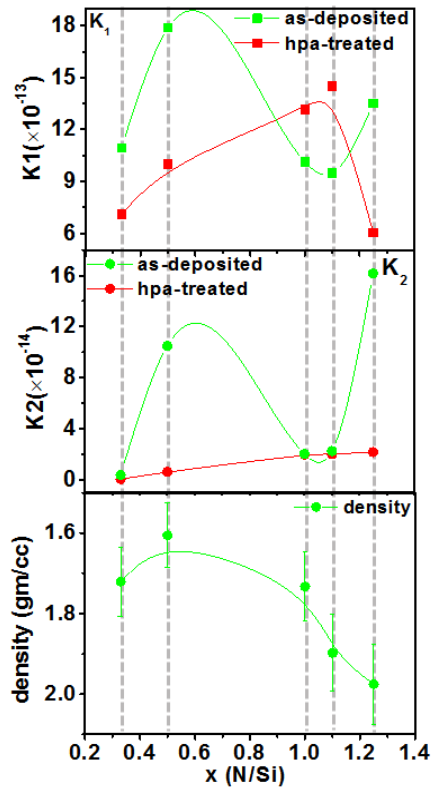

**Figure 5.** Decay rates (a) $K_1$ and (b) $K_2$ for the as deposited (green markers) and hydrogenated films (red markers) notice the order of magnitude difference in the values of $K_1$ and $K_2$, slow and fast depletion processes respectively. (c) The density determined from X-Ray Reflectivity critical angles.

## 4. Conclusions

Two types of diffusion processes have been found to determine hydrogen desorption from *a*-SiN$_x$:H films, namely, molecular and atomic diffusion. Molecular diffusion is a fast process and prevails in the initial stages of irradiation. The atomic diffusion is a slow process dominating in the high fluence regime. The critical hydrogen concentration (H$_{critical}$) determining the transition from one process to another was found to lie ~8±2 atoms/nm$^3$ and ~14±2 atoms/nm$^3$ for the as-deposited and HPA treated films respectively. The H$_{critical}$ values clearly show that the diffusion of atomic hydrogen is severely limited to within a ~1 nm$^3$ volume around the point of it creation. The improvement in the H$_{critical}$ values, upon HPA, are attributed to defect passivation and a consequent improvement in hydrogen radical mobility. Further, HPA was found to improve the hydrogen retention in *a*-SiN$_x$:H films by one order of magnitude for films with N/Si ratio ranging between 0.3 to 1.25. The confirmation of correlations between the decay rates and the stoichiometry observed here would require further detailed investigations.

**Acknowledgements**

Personnel at IUAC New Delhi are thanked for their support during the experiments. Institute of Physics Bhubaneswar is thanked for the postdoctoral fellowship. This work was partly funded by financial grant of IAEA [Grant No. 17254/R1].